# The hollow Gaussian beam propagation on curved surface based on matrix optics method

Weifeng Ding,[1] Zhaoying Wang[1, *]

[1] Zhejiang Province Key Laboratory of Quantum Technology and Device, Department of Physics, Zhejiang University, Hangzhou 310027, China

*E-mail address:* zhaoyingwang@zju.edu.cn

**Abstract**

In this paper, based on the Euler-Lagrange equation, an ABCD matrix is constructed out to study the paraxial transmission of light on a constant Gaussian curvature surface (CGCS), which is the first time to our knowledge. Then, by using the method of matrix optics, we extend the CGCS matrix to a general transfer matrix which is suitable for a gently varying curvature. As a beam propagation example, based on the Collins integral and the derived ABCD matrix elements, an analytical propagation formula for the hollow Gaussian beams (HGBs) on the CGCS is deduced. The propagation characteristics of HGBs on a CGCS are illustrated graphically in detail, mainly including the change of dark spots size and splitting rays. Besides its propagation periodicity and diffraction properties, a criterion for convergence and divergence of the spot size is proposed. The area of the dark region of the HGBs can easily be controlled by proper choice of the beam parameters and the shape of CGCS. In addition, we also study the special propagation properties of the hollow beam with a fractional order. Compared with propagation characteristics of HGBs in flat space, these novel propagation characteristics of HGBs on curved surface may further expand the application range of hollow beam.

## 1. Introduction

We know that under the general relativity theory, the gravitational field is described as space-time curvature, therefore, it is obvious that the study of the light beam propagation on curved space is meaningful for cosmology, such as the measurement of angular size of stars [1]. Moreover, the interaction between electromagnetic waves and curved space has a number of intriguing effects, such as Hawking radiation [2, 3] and Unruh effect [4]. On the other hand, it appears many analogical theories and experiments of the curved space since the work by Unruh [5]. For instance, use Bose-Einstein condensation system to simulate the black hole Gibbons Hawking effect [6, 7]. Recently, the moving dielectric medium [8, 9] and the nonlinear Schrödinger Newton system [10, 11] are used to demonstrate the gravitational field caused by the curved space. Thus, the study of the transmission characteristics of light on curved space is not only a generalization of flat space, but also a reference for a large number of analogical experiments.

Currently, two-dimensional surface is a common simplification model in the study of curved space, such as the two-dimensional Wolf effect of curved space [12], and other studies on geometric optics [13-15] and physical optics [16, 17] of two-dimensional manifolds. But in most cases [18, 19], due to the complexity of curved surface algebra and differential geometry which need a lot of numerical calculations, sometimes there are singularities, greatly reducing the generality of the analytical expression and thus reducing its physical significance. Moreover, most of the research work is based on the transmission along the generatrix, this paper mainly studies the equatorial transmission. Due to the effective generalization of WKB approximation from Euclidean space to curved surface [20], we need to further study whether other optical transformation methods can be effectively generalized. Collins' work [21] shows that ground on the content of simple geometric optics, we can study the properties of physical optics under complex optical transformations. Therefore, in this paper, we hope to utilize the tool -- ABCD matrix and Collin's formula to describe the curved surface optics, which is undoubtedly universal to explore the light propagation characteristics on the curved surface under the paraxial approximation.

Furthermore, since Lin's work [22], the propagation properties of hollow Gaussian beams(HGB) have been gradually revealed, such as its generation [23], far-field structure [24] and the propagation through misaligned optical systems [25]. Hollow Gaussian beam originates from simple Gaussian light and is a typical representative of all other hollow light. Founded on the HGB, a large number of other hollow beams [26-28], including vortex beams, are derived, which have a wide range of applications.

In our previous work [29], the eikonal function of light rays on a CGCS was derived based on the wave equation. In this paper, we will obtain the CGCS transfer matrix from the Euler-Lagrange equation and further extend it to a general surface by using the language of matrix optics. We prove that this method has good universality. Additionally, we study the related propagation properties of a hollow beam on the CGCS. Due to the periodic nature of CGCS transmission, some properties of HGB can even be applied to the far field. We hope that this research can provide ideas for the beam propagation on other surfaces.

## 2. The matrix optical method of curved surface

In this paper, we firstly consider the transfer matrix of the beam propagation on a CGCS and then generalize it to an ordinary curved surface in paraxial approximation. The CGCS model, with an assumption called surface of revolution, will be formed by the rotation of certain radius parameter $\rho(h)$, which is the distance between one point on surface and the axis of symmetry.

For a two-dimensional surface which is a sub-manifold imbedded into three-dimensional space, we define the expression of CGCS as $\rho(h) = r_0 \cos(h/r)$, where $r_0$ and $r$ are transverse parameter. During the propagation of beam, there is an entrance point and an exit point on this surface, the position $h_1$, $h_2$ and transmission direction $h_1'$, $h_2'$ of the lights as shown in Fig. 1 (a). According to Fermat's principle on the surface [19], we know that light travels along geodesics, then we can calculate the propagation path by using Lagrange's method. First of all, the line element of CGCS: $ds^2 = dh^2 + \rho^2(h)d\theta^2$ can be constructed as a D'Alembert action $s = \int_{h_1}^{h_2} \sqrt{1 + \rho^2(h)\dot{\theta}^2} dh$, the Euler-Lagrange equation is written as follows:

$$\frac{d}{dh}\frac{\partial L}{\partial \dot{\theta}} - \frac{\partial L}{\partial \theta} = 0. \quad (1)$$

Where $\theta$ is the angle of rotation around the axis of symmetry between the incident point and the exit point, $h$ is the arc length from the alternative point towards the maximum rotational circuit, $h' = dh/(r_0 d\theta)$ represents the transmission direction and the Lagrangian $L = \sqrt{1 + \rho^2(h)\dot{\theta}^2}$. According to Eq. (1), we can obtain the propagation path, which satisfy the following expression:

$$\tan(h/r) = \sin\left(\frac{r_0}{r}(\theta - c_1)\right)c_2. \quad (2)$$

Here $c_1$, $c_2$ are the integral constants. We take the derivative of both sides with respect to $\theta$,

$$\frac{1}{r}\sec^2(h/r)\frac{dh}{d\theta} = \frac{r_0}{r}\cos\left(\frac{r_0}{r}(\theta - c_1)\right)c_2. \quad (3)$$

According to our coordinate definition, let's assume the initial conditions are $h(0) = h_1$, $h'(0) = h_1'$ and the final condition are $h(\theta) = h_2$, $h'(\theta) = h_2'$, the relationship between the entrance point and an exit point can be derived:

$$\tan\frac{h_2}{r}\cos^2\frac{h_1}{r} = h_1'\sin\frac{r_0}{r}\theta + \frac{1}{2}\sin\frac{2h_1}{r}\cos\frac{r_0}{r}\theta,$$
$$h_2' = h_1'\cos\frac{r_0}{r}\theta - \frac{1}{2}\sin\frac{2h_1}{r}\sin\frac{r_0}{r}\theta. \quad (4)$$

Perform the paraxial approximation and ignore the higher-order small terms, we can get the linear relationships: $\sin(h_{1,2}/r) \approx \tan(h_{1,2}/r) \approx h_{1,2}/r$, $\cos(h_{1,2}/r) \approx 1$. That is to say, the transfer function of beam is the homogeneous expression of $h_1$ and $h_2$. So, in order to better describe the propagation characteristics of light, we introduce the method of matrix optics into the curved surface system in the following:

$$\begin{pmatrix} h_2 \\ h_2' \end{pmatrix} = \begin{pmatrix} A & B \\ C & D \end{pmatrix}\begin{pmatrix} h_1 \\ h_1' \end{pmatrix} = \begin{pmatrix} \cos\frac{r_0\theta}{r} & r\sin\frac{r_0\theta}{r} \\ -\frac{1}{r}\sin\frac{r_0\theta}{r} & \cos\frac{r_0\theta}{r} \end{pmatrix}\begin{pmatrix} h_1 \\ h_1' \end{pmatrix}. \quad (5)$$

From Eq. (5), we can also deduce that the eikonal function of light rays between the incident point and the exit point had the following form as Eq. (6), which is consist to our previous work [29]. The method we employ here is simpler, has fewer parameters, and avoids the tedious algebra of differential geometry.

$$s \approx s_0 + \frac{1}{2r}\left(h_1^2 \cot\frac{r_0}{r}\theta - 2h_1h_2\csc\frac{r_0}{r}\theta + h_2^2\cot\frac{r_0}{r}\theta\right). \quad (6)$$

The introduction of matrix optics can facilitate the calculation of paraxial optics. In particular, the matrix optics greatly reduces the computational complexity in engineering, where the curved surface may be combined with several optical devices. The significant advantage is that the transformation matrix of the complex system can be simply acquired by multiplied the matrix of the optical devices in order.

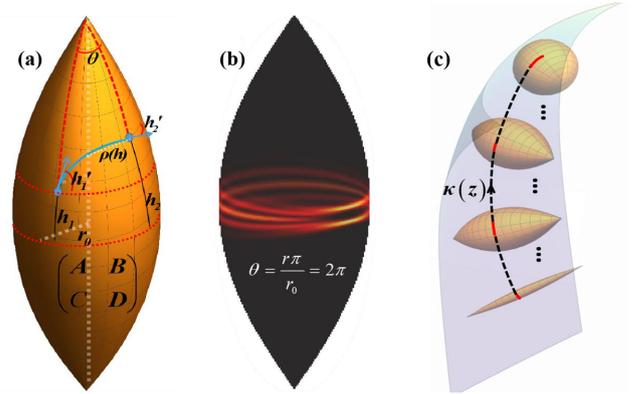

Fig. 1 (a) Definition of the surface coordinates and parameters. The blue line represents a light ray on the surface. (b) HGB propagation diagram on the surface, which shows the simulation of the light intensity distribution when $n=3$ and $r/r_0 = 2$. (c) A general surface can be thought of as a combination of CGCS.

The derivation of ABCD matrix is completely dependent on geometrical optics, which means that the ABCD matrix can be derived by paraxial approximation of the geodesic equation on the surface in addition to using the definition of the ray equation. Here, our study on the surface of constant Gaussian curvature can be extended to an ordinary curved surface. To make the matrix more general, we can rewrite Eq. (5) by using the transmission distance $z = r_0\theta$ and the Gaussian curvature $\kappa = r^{-2}$.

$$M(z) = \begin{pmatrix} \cos(\sqrt{\kappa}z) & \frac{1}{\sqrt{\kappa}}\sin(\sqrt{\kappa}z) \\ -\sqrt{\kappa}\sin(\sqrt{\kappa}z) & \cos(\sqrt{\kappa}z) \end{pmatrix}. \quad (7)$$

$z$ axis is usually the geodesic of surface. For a general curved surface, $\kappa$ is not constant, but varies with the change of $z$, set as $\kappa(z)$. According to the idea of the element method, as shown in Fig. 1 (c), a general curved surface is divided into more enough CGCS segments. The total equivalent matrix is going to be the product of all of these matrices of CGCS segments:

$$M(\kappa(z),z) = \lim_{n\to\infty}\left[\prod_{m=n}^{1} M\left(\kappa(z_m),\frac{z}{n}\right)\right] = \lim_{n\to\infty}\prod_{m=n}^{1}\begin{pmatrix} \cos\left(\frac{\sqrt{\kappa(z_m)}z}{n}\right) & \frac{1}{\sqrt{\kappa(z_m)}}\sin\left(\frac{\sqrt{\kappa(z_m)}z}{n}\right) \\ -\sqrt{\kappa(z_m)}\sin\left(\frac{\sqrt{\kappa(z_m)}z}{n}\right) & \cos\left(\frac{\sqrt{\kappa(z_m)}z}{n}\right) \end{pmatrix}. \qquad (8)$$

The symbol of $\prod_{m=n}^{1}$ means multiplying matrices in reverse order, $n$ is the number of segments. Using the trigonometric properties and gently varying curvature approximation ($\kappa(z_m)\approx\kappa(z_{m+1})$), Eq. (8) can be converted to the following form:

$$M(\kappa(z),z) \approx \lim_{n\to\infty}\begin{pmatrix} \cos\left(\sum_{m=1}^{n}\frac{\sqrt{\kappa(z_m)}z}{n}\right) & \frac{n}{\sum_{m=1}^{n}\sqrt{\kappa(z_m)}}\sin\left(\sum_{m=1}^{n}\frac{\sqrt{\kappa(z_m)}z}{n}\right) \\ -\sum_{m=1}^{n}\frac{\sqrt{\kappa(z_m)}}{n}\sin\left(\sum_{m=1}^{n}\frac{\sqrt{\kappa(z_m)}z}{n}\right) & \cos\left(\sum_{m=1}^{n}\frac{\sqrt{\kappa(z_m)}z}{n}\right) \end{pmatrix} \approx \begin{pmatrix} \cos\left(\int_{z_1}^{z_2}\sqrt{\kappa(z)}dz\right) & \frac{1}{\sqrt{-\bar\kappa}}\sin\left(\int_{z_1}^{z_2}\sqrt{\kappa(z)}dz\right) \\ -\sqrt{\bar\kappa}\sin\left(\int_{z_1}^{z_2}\sqrt{\kappa(z)}dz\right) & \cos\left(\int_{z_1}^{z_2}\sqrt{\kappa(z)}dz\right) \end{pmatrix}. \qquad (9)$$

Where $\bar\kappa = \int_{z_1}^{z_2}\kappa(z)dz/(z_2-z_1)$ from and says the average Gaussian curvature. Eq. (9) is worth noting that the matrix is valid only for the surface when Gaussian curvature change is not too big.

As we have the transformation matrix of CGCS system, we can study the beam propagation properties passing through this kind of optical system. Combining with the ABCD matrix and the Collins formula, the output light field $E_2(h_2)$ can be obtained after the input light field $E_1(h_1)$ passing through the CGCS system, as shown in Eq. (10):

$$E_2(h_2) = \left(-\frac{ik}{2\pi B}\right)^{1/2} e^{ikL_0} \int_{-\infty}^{\infty} E_1(h_1)\exp\left[\frac{ik}{2B}\left(Ah_1^2 - 2h_1h_2 + Dh_2^2\right)\right]dh_1. \qquad (10)$$

## 3. Propagation of hollow Gaussian beams on curved surfaces

As was mentioned at the beginning, in this paper, the derived ABCD matrix will be utilized to study a special case – the propagation of hollow Gaussian beam on CGCS. We can not only visually see the propagation properties of the split light, but also study its hollowness, a property common to many other hollow beams (including vortex beams). In general, we set the electric field of the initial hollow Gaussian beam simply as follows[22]:

$$E_1(h_1) = \left(\frac{h_1^2}{\sigma^2}\right)^n e^{-\frac{h_1^2}{\sigma^2}}. \qquad (11)$$

In which, $n$ represents the order of the HGB and the initial transverse spot size is $w_h = 2\sqrt{n}\sigma$. Using the method depicted in the first part, the Eq. (11) is substituted into the Collins' formula Eq. (8), then an analytic solution is deduced:

$$E_2 = \frac{1}{\sqrt{2\pi}} e^{\frac{iDkh_2^2}{2B}+ikr_0\theta}\sqrt{\frac{k}{iB}}\left(\frac{1}{\sigma^2}-\frac{iAk}{2B}\right)^{-\frac{1}{2}-n}\left(\frac{1}{\sigma^2}\right)^n \\ \times \Gamma\left(\frac{1}{2}+n\right)HF1\left(\frac{1}{2}+n,\frac{1}{2},-\frac{h_2^2 k^2\sigma^2}{4B^2-2iABk\sigma^2}\right). \qquad (12)$$

Where $\Gamma(x)$ denotes to the gamma function. $HF1(a,b,x)$ represents the Cumor confluence hypergeometric function and the first few terms of the expansion are

$$HF1(a,b,x) = 1 + \frac{ax}{b} + \frac{a(1+a)x^2}{2b(1+b)} + \frac{a(1+a)(2+a)x^3}{6b(1+b)(2+b)} + O[x^4]. \qquad (13)$$

For purpose of facilitate the discussion of the propagation properties of hollow beam, the curved surface is flattened into a plane surface and a two-period image can be drawn, as shown in Fig. 2 (a). Horizontal coordinate of Fig. 2 (a) is the angle $\theta$ of revolution for transmission and its vertical coordinate is $h$. Figure 2 (a) is the light intensity flattened maps of Gaussian beam and hollow Gaussian beam during the propagation on the curved surface. We explore that the hollow Gaussian beam can keep the hollow characteristic in the near field, but becomes no longer "hollow" in the far field, the area of dark region depends on $n$. For contrast, the propagation of hollow beams in a flat space is presented in Fig. 2 (b).

In the following, we focus on the beam divergence and convergence properties during transmission. Before that, the study of Gaussian beam (i.e., $n = 0$) can serve as a reference for us, the transverse waist width of the Gaussian beam is calculated as follows:

$$w_{h,n=0} = \sigma\sqrt{\frac{4B^2}{k^2\sigma^4}+A^2}, \qquad (14)$$

the change of waist width equals to

$$\frac{dw^2_{h,n=0}}{d\theta} = \frac{r_0 \sigma^2}{r} \sin\left(\frac{2r_0\theta}{r}\right)\left[\left(\frac{2r}{k\sigma^2}\right)^2 - 1\right]. \quad (15)$$

From Eq. (15), we explore that the derivative is a periodic function with period $\theta = \pi r / r_0$, and the first half of the period is positive or negative depending on the parameter $\gamma_d = 2r/(k\sigma^2)$, which is called the divergence coefficient in this paper. This coefficient is closely related to the original beam size and the shape of surface. When $\gamma_d < 1$, the beam converges and then diverges, when $\gamma_d > 1$, the beam first diverges and then converges after emission, and in the case of $\gamma_d = 1$, the variation of Gaussian beam spot size vanishes. The last two transmission features are demonstrated in the first raw plot of Fig. 2 (a). But as the order $n$ increases, it does not work that way. The second-order moment is used to describe the overall mean spot width:

$$w_{h,n}^2 = \frac{\int_{-\infty}^{\infty} h_2^2 |E_2|^2 dh_2}{\int_{-\infty}^{\infty} |E_2|^2 dh_2} = \frac{4n+1}{4}\left[1 + \sin^2\left(\frac{r_0\theta}{r}\right)\left(\frac{8n-1}{16n^2-1}\gamma_d^2 - 1\right)\right]. \quad (16)$$

So, we can use the change in average waist width as a criterion for overall convergence and divergence. According to Eq. (16), now the divergence coefficient changes from $\gamma_d$ to $\sqrt{(8n-1)/(16n^2-1)} \times \gamma_d$.

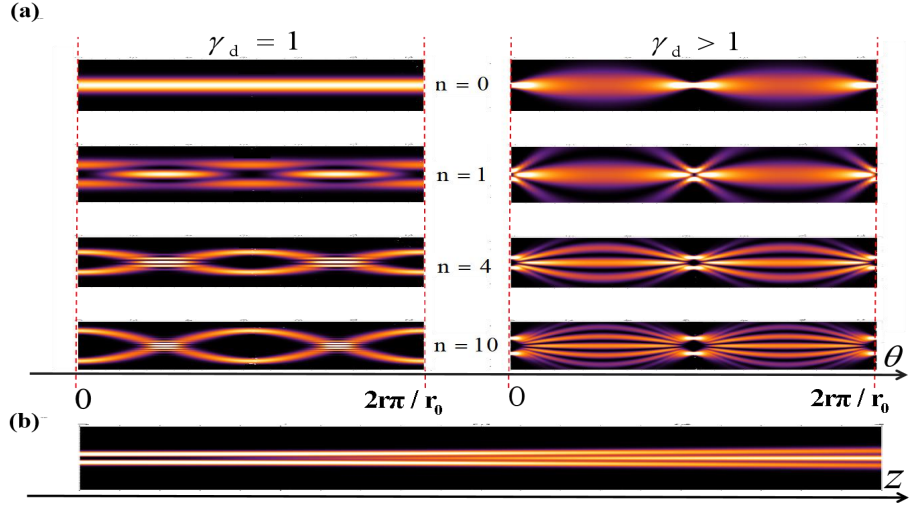

**Fig. 2.** Propagation of the $n$-order HGBs on curved surfaces(a) and flat surfaces(b). On the curved surfaces, we calculated the propagation of a HGB of different order $n$ ($n$ = 0,1,4,10). Each plot holds two periods. As observed, with the increase of $n$, the maximum number of splitting rays increased in the process of transmission. Transverse comparison shows that $\gamma_d$ significantly affects the dark spot size.

Compared with Fig. 2 (a) and (b), we also can discover that the curved surface accelerates the near-field and the far-field conversion of light transmission, i.e., the fast conversion between the Fresnel diffraction region and the Fraunhofer diffraction region within half a transmission period on the curved surface. So, we think that the curved space of creating far-field conditions over short distances can be considered for the manufacture of optical elements.

In the case of Gaussian beam, the light intensity distribution along $h$ direction has only one single peak, while for the propagation process of HGB, the transverse light intensity develops from bimodal image to multiple peaks. The maximum number of peaks is equal to $2n+1$, which can also be proved analytically by Eq. (12). In order to obtain the transverse positions of each peak, we can calculate the extreme value of light intensity during the half period based on Eq. (12), we demonstrate that their positions satisfy the following expression.

$$h_2 \sum_{i=0}^{n}\sum_{j=0}^{n}\left(-\frac{h_2^2 k^2 \sigma^2}{r^2}\right)^{i+j} \frac{(n!)^2}{(n-i)!(n-j)!(2i)!(2j+1)!} = 0. \quad (17)$$

From Eq (17), we can see that, except for the zero point at infinity, $h_2$ has $(4n+1)$th order at most, that is, there are $4n+1$ solutions corresponding to $4n+1$ extremum points of light intensity, including $2n+1$ peaks and $2n$ valleys, alternately.

Considering the Gaussian beam, when the light diverges, the intensity on the propagation axis becomes weaker, and vice versa, but for the hollow light, the situation is more complicated. Its light intensity on the propagation axial is as follows:

$$|E_2'|^2\Big|_{h_2=0} = \left(\frac{(2n-1)!}{2^{2n-1}(n-1)!}\right)^2 \frac{\gamma_d^{2n}\left(\gamma_d^2 + \cot^2\frac{r_0\theta}{r}\right)^{-n-1/2}}{\sin\frac{r_0\theta}{r}}. \quad (18)$$

According to Eq. (18), we divide the intensity shapes on the propagation axis into two types as revealed in Fig. 3: A type and B type. There is only one maximum-intensity between two minimum-intensity positions for A type, but there are two maximum-intensity positions for B type. Surprisingly, the critical boundary conditions for these two types are very simple, that is, whether $\gamma_d$ is greater than $\sqrt{2n+1}$. If the light intensity distribution is A type, otherwise, it is B type. It means besides the curvature of the surface and the Rayleigh distance of the beam; the order of the hollow beam can directly affect its hollowness. Next, we will discuss the problem of hollowness, which will involve the definition of the length of dark spot. It is shown that the maximum intensity of type A and the second minimum intensity of type B are both at the half period position with $\theta = \pi r / 2r_0$. The longitudinal length of the dark spot on the propagation axis $w_\theta$ is defined as the distance from the center of the dark spot to half of the maximum light intensity (type A), or to the first maximum intensity (type B), as shown in Fig. 3.

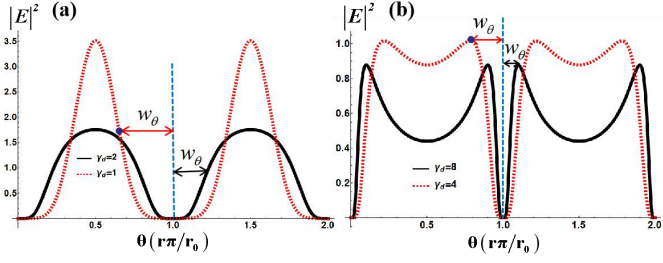

**Fig. 3**. Two types of light intensity distribution. The transverse coordinate represents the angular distance on the propagation axis, the longitudinal coordinate represents the light intensity. A-type (a) and B-type (b) are unimodal and bimodal structures within one period, respectively.

For type A, $w_\theta$ has no analytic solution and satisfies the following equation:

$$\frac{\left(\gamma_d^2 + \cot^2 \frac{r_0 w_\theta}{r}\right)^{-n-1/2}}{\sin \frac{r_0 w_\theta}{r}} = \frac{\gamma_d^{-2n-1}}{2}, \quad \gamma_d \leq \sqrt{2n+1}. \quad (19)$$

While based on Eq. (18), We can simply obtain the size of the dark spot of type B by calculating the maximum value of the light intensity as:

$$w_\theta = \frac{r}{r_0}\arcsin\left(\sqrt{\frac{2n}{\gamma_d^2 - 1}}\right), \quad \gamma_d > \sqrt{2n+1}. \quad (20)$$

Here, we mainly focus on the beam propagation of type B. With the increase of $r$, the longitudinal size of the dark spot becomes smaller, but tends to a constant value $r_0 w_\theta = \sqrt{2n}k\sigma^2/2$, which is consistent with the result of flat space[22]. It is also obvious that the square of the longitudinal widths is linear with $n$ in flat space, while on curved surface, its hollow region will keep longer, and it will grow faster than flat space as $n$ increases. However, the transverse size of our dark spot is consistent with that of the flat space, which indicates that our treatment of the curved surface follows such a phenomenon: the paraxial approximation ignores the surface property of light perpendicular to the propagation axis, but focuses on the surface property along the propagation axis.

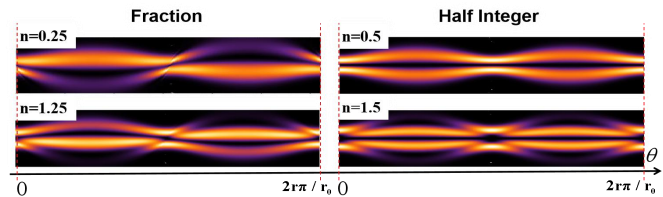

**Fig. 4**. Fractional order beam transmission images on the curved surfaces. The intensity of half-integer-order beams on the axis remain dark during its propagation, other fractional orders can cause asymmetry of transverse light intensity distribution.

All the above work is based on the fact that the beam order $n$ is an integer, but $n$ can also be a fraction. Next, we focus on the optical transmission characteristics when $n$ is a fraction. Figure 4 displays the examples when the value of $n$ is fraction and half-integer. As we can see, the most intuitive one is that when $n$ is half integer, HGB remains hollow intensity along the propagation axis all the time, which is as same as the beam propagation on flat space. Besides, for other general fractions, the beam fringes in adjacent periods are distorted, causing periodic changes in the axial light intensity. Here, we can use the statistical skewness to describe the overall distortion of the light field distribution in two adjacent periods. For convenience, we can take the half period light field distribution as the skew analysis, and the results are presented in Fig. 5. Here, we take the position of half period within two adjacent periods as an index to describe the overall skewness. We find that, with the change of $n$, the skewness oscillates and reaches zero when n is an integer and a half integer. Moreover, the variation of skewness is complementary for two adjacent periods, as shown by the red solid line and the black dotted line.

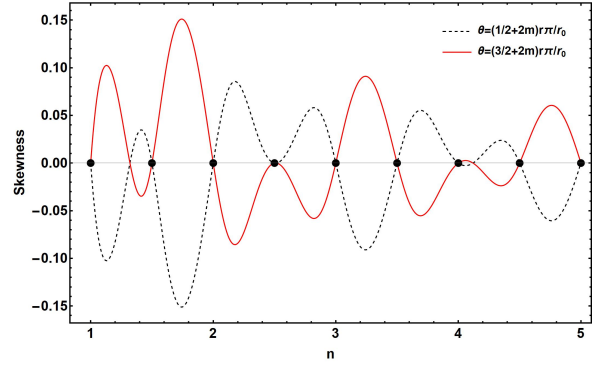

**Fig. 5**. Skewness changes with respect to $n$. The parameter $m$ denotes to an arbitrary integer.

The Fig. 5 indicates that the centroid of the beam deviates from the transmission axis at most values of $n$, showing the novel property of the self-bending of the fractional hollow beam transmitting along the equator on the curved surface. When $n$ is an arbitrary number, we can also get a more general magnitude of the light intensity on the axis. The light intensity on the propagation axial is as follows:

$$\left|E_2'\right|^2_{h_2=0} = \frac{\cos^2(n\pi)\Gamma^2\left(\frac{1+2n}{2}\right)\gamma_d^{2n}\left(\gamma_d^2 + \cot^2 \frac{r_0\theta}{r}\right)^{-n-1/2}}{\pi \sin \frac{r_0\theta}{r}}. \quad (21)$$

From Eq. (21), it can be seen that, due to the existence of the factor $\cos^2 n\pi$, the light intensity on the propagation axis oscillates with $n$. When $n$ is half integer, the light intensity on the propagation axis is zero everywhere, which is very interesting for the optical trapping or the optical guiding.

## 4. Conclusion

In conclusion, we introduce a ABCD matrix of the CGCS based on the Euler-Lagrange equation. Then, we extend the results to a general surface by using the properties of ABCD matrix. As the focus of the whole manuscript, we derive an analytical propagation formula for HGBs on CGCS by using the Collins integral and plot the transmission properties of HGB. We find the periodic divergence and convergence of light and the light is split into some rays in the process of transmission. The longitudinal dark area depends on the order $n$, the original beam size and the shape of curved surface, which is almost different from that of flat space. The transverse dark size is totally consistent with that of flat space. Furthermore, we discuss the optical transmission of HGB when *n* is a fraction. The ABCD matrix has been shown to be an ideal and convenient model with which to describe beam propagation on CGCS.

In addition, our matrix processing method is not limited to the propagation on the selected axis. Since the misaligned optical system can be further described by using a $4\times 4$ tensor matrix, in the future, we can also deal with the cases of eccentricity or rotation of HGB on curved surface[25].

## Declaration of Competing Interest

The authors declare that they have no known competing financial interests or personal relationships that could have appeared to influence the work reported in this paper.

## Acknowledgements
National Key Research and Development Program of China (2017YFC0601602).